\documentclass[twocolumn]{aa} 

\def\rmit#1{{\it #1}}              
\def\specchar#1{{\sc #1}}

\def\ie{\rmit{i.e.}}
\def\eg{\rmit{e.g.}}

\def\arcsec{\hbox{$^{\prime\prime}$}}
\def\FeI{\mbox{Fe\,\specchar{i}}}

\def\BaII{\mbox{Ba\,\specchar{ii}}}
\def\CaIIK{\mbox{Ca\,\specchar{ii}\,\,K}}       
\def\Halpha{\mbox{H$\alpha$}}

\usepackage{natbib}
\usepackage{graphicx}
\begin{document}
%
   \title{Solar granulation from photosphere to low chromosphere
   observed in \BaII\ 4554 \AA\ line}

   \author{R. Kostik\inst{1}, E. Khomenko\inst{2,1} and N.
   Shchukina\inst{1}}

   \institute{Main Astronomical Observatory, NAS, 03680, Kyiv, Ukraine\\ \email{kostik@mao.kiev.ua}
         \and Instituto de Astrof\'{\i}sica de Canarias, 38205 La Laguna, Tenerife, Spain\\
             }

   \date{Received XXX, 2008; accepted xxx, 2008}

\abstract
   {}  
{The purpose of this paper is to characterize the statistical
properties of solar granulation in the photosphere and low
chromosphere up to 650 km.}
{We use velocity and intensity variations obtained at different
atmospheric heights from observations in \BaII\ 4554 \AA. The
observations were done during good seeing conditions at the VTT at
the Observatorio del Teide on Tenerife. The line core forms rather
high in the atmosphere and allows granulation properties to be
studied at heights that have been not accessed before in similar
studies. In addition, we analyze the synthetic profiles of the
\BaII\ 4554 \AA\ line  by the same method computed taking NLTE
effects into account in the 3D hydrodynamical model atmosphere. }
{We suggest a 16-column model of solar granulation depending on
the direction of motion and on the intensity contrast measured in
the continuum and in the uppermost layer. We calculate the heights
of intensity contrast sign reversal and velocity sign reversal. We
show that both parameters depend strongly on the granulation
velocity and intensity at the bottom photosphere. The larger the
two parameters, the higher the reversal takes place in the
atmosphere. On average, this happens at about 200--300 km. We
suggest that this number also depends on the line depth of the
spectral line used in observations. Despite the intensity and
velocity reversal, about 40\% of the column structure of
granulation is preserved up to heights around 650 km.}
{}



\authorrunning{Kostik et al.}
\titlerunning{Solar granulation from photosphere to low chromosphere}

   \maketitle
%
\section{Introduction}

Convective motions in the solar atmosphere have been studied in
detail during the past decades, using different types of
observations, their analysis, and interpretation. The summary of
the investigations before 2000 can be found in
\citet{Espagnet+Muller+Roudier+Mein+Mein1995} and
\citet{Brandt2000}.
According to the final results, the photosphere can be divided
into two regions with different physical properties with the
border between them lying at heights about 170 km above the
continuum formation height \citep[see][for
details]{Espagnet+Muller+Roudier+Mein+Mein1995}.
This conclusion was reached based on the amplitudes of intensity
and velocity fluctuations first decreasing with height, reaching
minimum at about 170 km and then starting to increase again.
The intensities and velocities at heights above and below 170 km
do not seem to be correlated with each other.
On the other hand, \citet{Espagnet+Muller+Roudier+Mein+Mein1995}
came to the conclusion, based on their own investigations, that
the photosphere can be divided into these two regions only in
intensity fluctuations, since the vertical convective velocity
structure is preserved until heights as high as 500 km.
A similar result has also been obtained by
\citet{Salucci+etal1994}.

During the past few years significant efforts have been made to
determine the precise height where the inversion of intensity
contrast (or temperature) takes place. The results by different
authors diverge significantly in this respect.
Values as low as 60 km are found by \citet{Kneer+etal1980} from
spectrograms of the Mg\,{\sc i} $\rm b_2$ line taken at the quiet
sun disk center. While \citet{Bendlin+Volkmer1993} give values as
high as 270 km from 2D spectral images obtained with tuning a
Fabry-Perot interferometer in the \FeI\ 6303 \AA\ line in a
moderately active region near disk center.
Different values are also given in the more recent literature,
\eg\ 250 km \citep{Kostyk+Shchukina2004}, 200 km
\citep{Janssen+Gauzzi2006}, 170 km \citet{Puschmann+etal2005}.
In the last paper, the authors perform a careful analysis and NLTE
inversion of the observed spectral profiles of the several
photospheric \FeI\ lines and present the resulting temperature and
velocity stratifications as functions of both optical depth and
geometrical height. In addition to the height of inversion of the
intensity contrast (temperature), they also report that only the
large-scale structures (about 4\arcsec) penetrate to higher layers
without the lost of correlation though the whole atmosphere.
Similar results were also reported by \citet{Puschmann+etal2003},
where the authors find that only the intensity structures with
sizes over 2\arcsec\ at heights about 435 km are still connected
with those at the continuum level.

In all the papers reported above
\citep[except][]{Kostyk+Shchukina2004}, the vertical velocities
were observed in phase though the whole photosphere penetrating
into the highest layers under study. Only
\citet{Kostyk+Shchukina2004} have obtained evidence of the
velocity reversal at heights above 490 km for the elements with
the highest contrast in their data (above 6-9\%).

The spectral lines used in previous studies gave the possibility
of analyzing statistical properties of solar granulation mainly in
photospheric layers below the temperature minimum.
The aim of the present paper is to extend this analysis to the
higher layers taking advantage of spectral observations in the
\BaII\ 4554 \AA\ line.
According to \citet{Olshevsky+etal2007} and
\citet{Sutterlin+etal2001}, the core of this line forms rather
high at about 700 km. It makes the line particularly useful for
studying the vertical structure of the solar atmosphere, since we
can follow this structure to much greater heights than in the
previous analysis.
Recent theoretical investigation \citep[][hereafter referred to as
Paper I]{Shchukina+etal2009} based on NLTE modeling the \BaII\
4554 \AA\ line in a three-dimensional (3D) hydrodynamical model
have shown that this line is a valuable tool for the Doppler
diagnostics along the whole photosphere and lower chromosphere.

In the present study we use theoretical understanding attained in
Paper I for interpreting the new high-resolution spectral
observations in the \BaII\ 4554 \AA\ line.
We investigate the column structure of granulation up to the
heights above the solar temperature minimum,
derive  the height of the intensity contrast sign reversal and the
change of sign of the vertical motions, and propose a 16-column
model of granulation, depending on the direction of motion and
intensity in the low and high photospheres.
The results of observations are compared with the NLTE
calculations of the \BaII\ 4554 \AA\ line profiles in the 3D HD
model of solar convection by
\citet{Asplund+Nordlund+Trampedach+AllendePrieto+Stein2000b}.


\section{Observations}

Observations were obtained in July 21, 2004 at the 70-cm German
Vacuum Tower Telescope \citep[VTT, see][for
description]{Schroter+Soltau+Wiehr1985} at the Observatorio del
Teide in Tenerife during the good seeing conditions. The
observations were done with Echelle spectrograph.
The spectral region was centered on the \BaII\ 4554 \AA\ line.
Using the narrow-band \CaIIK\ and \Halpha\ slit-jaw images, we
selected a quiet region free of magnetic activity close to the
solar disk center.
Spectral images were recorded with a CCD camera with 1024 by 1024
pixels. The size of the images in spectral direction was 4.1 \AA.
The width of the spectrograph entrance slit was 100 $\mu$m
(equivalent to 0.\arcsec46). One pixel corresponds to 0.\arcsec087
on the solar surface. This way, the field of view of the telescope
was 0.\arcsec46 by 89\arcsec. However, after the reduction, only
about 730 pixels were found to be useful for the analysis,
equivalent to the 63.\arcsec5 of the solar surface.
The time series of \BaII\ 4554 spectral images was obtained with a
fixed slit position of the total duration of 69.9 minutes with an
interval of 7 seconds. The exposure time was 1 second, and a total
of 600 exposures were taken.
The image of the observed area was stabilized using the adaptive
optics (AO) system guiding on solar granulation. The average
Fried's parameter, as given by AO, was fluctuating around
$R_0=7-8$.
The spatial resolution of observations limited by the image motion
on the spectrograph slit due to earth atmospheric turbulence was
estimated to be about 0.\arcsec3--0.\arcsec7.


\section{Reduction of observations and calculation of line parameters}

Following the standard procedure, all 600 spectral images were
reduced for flatfield \citep[\eg][]{Kiselman1994}.
The flat field images were recorded immediately after the time
series of \BaII\ spectra in the same wavelength at the solar disk
center. The spatial structure of the solar surface was removed
from the flatfield images by rapid motions of the telescope scan
mirror. The reduction for flatfield included the following steps:
\begin{itemize}

\item removing the dark current out of the spectral images and the
flatfield images,

\item correction of all the images for the inclination and
distortion of the spectrograph sit with respect to the pixel
direction of the CCD,

\item removing the solar spectral lines from flatfield images,

\item splitting the flatfield images into the two components. Here
we need additional explanations. Both CCD camera and the
spectrograph slit have some dust on them. The dust at the CCD
camera appears as dark dots on the spectral images, while the dust
at the slit appears as dark horizontal lines, similar to
intergranular lanes. As the dust position on the slit can change
with time, the position of these dark lines on the spectral images
and on the flatfield images is not always the same. Thus, we had
to split the flatfield images into the images containing only a
CCD dust component (matrix flatfield) and those containing only
the slit dust component (slit flatfield).

\item removing the matrix flatfield from spectral images. This can
be done automatically. Removing the slit flatfield requires
individual control of all the spectral images, since the positions
of the dark lines do not always coincide. In some cases we had to
recalculate the slit flatfield images by shifting them in a
spatial direction by a fraction of a pixel.

\end{itemize}

The next step in the data reduction included calculating of the
intensity and velocity variations at different positions of the
\BaII\ line profiles. For that we used a $\lambda$-meter
technique. The description of this method can be found in
\citet{Stebbins+Goode1987} and
\citet[][Paper I]{Shchukina+etal2009}.

\begin{figure}
\centering
\includegraphics[width=9cm]{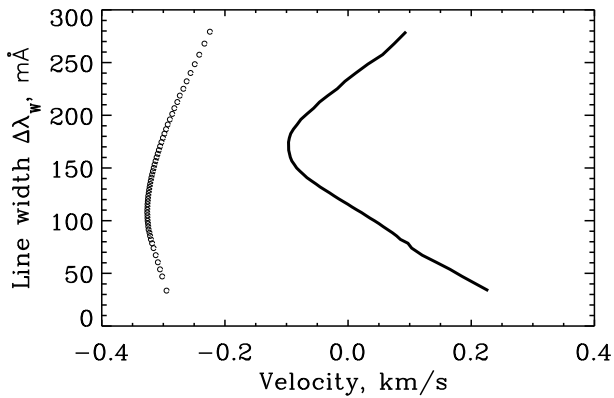}
\caption{Bisector due to hyperfine and isotopic shift calculated
in the MACKKL model \citep{malt:etal:1986} (small open circles),
and bisector of the observed profile, average over space and time,
caused by the nonthermal velocity field (solid line). }
\label{fig:asym}
\end{figure}

\begin{figure*}
\centering
\includegraphics[width=9cm]{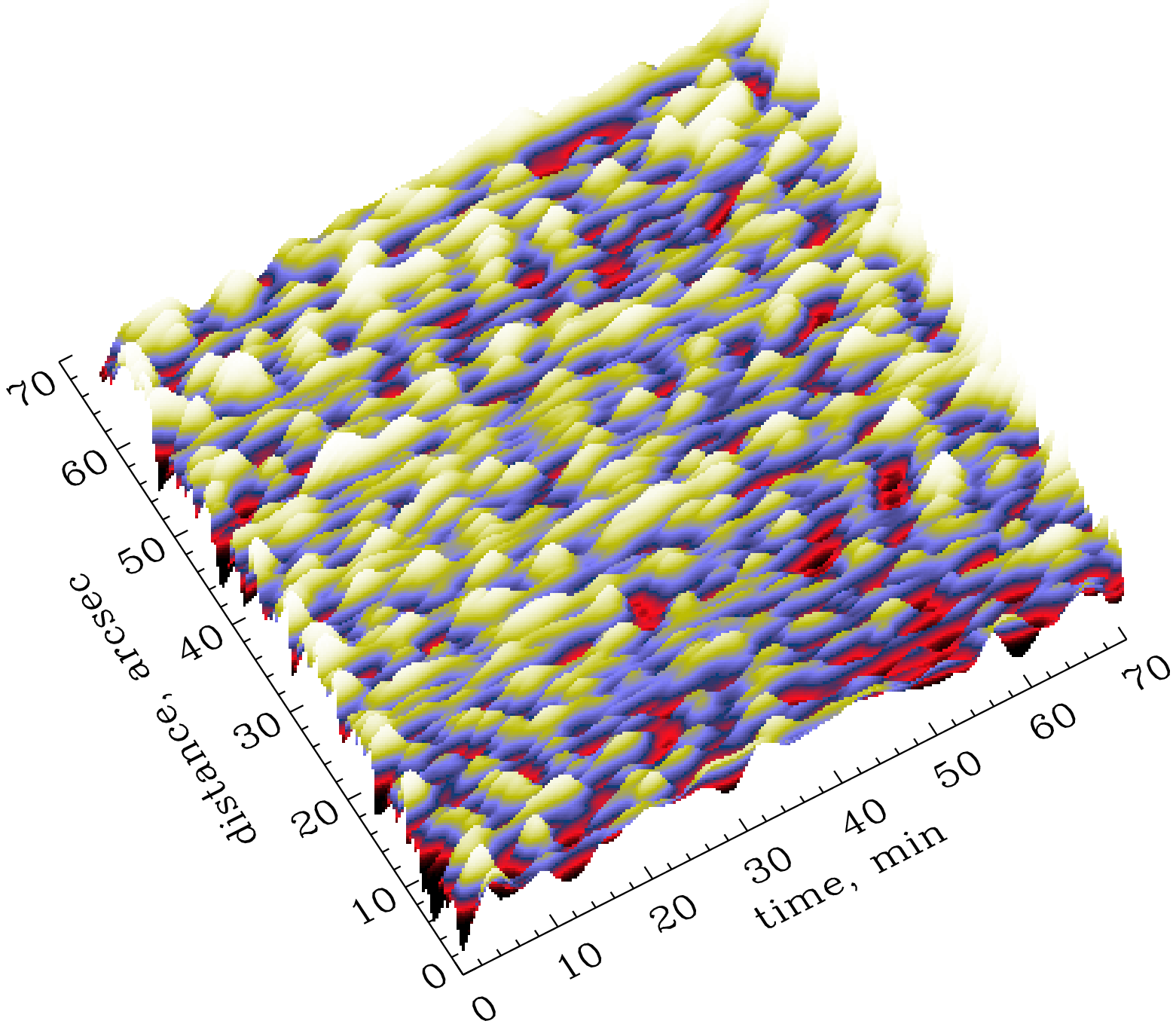}
\includegraphics[width=9cm]{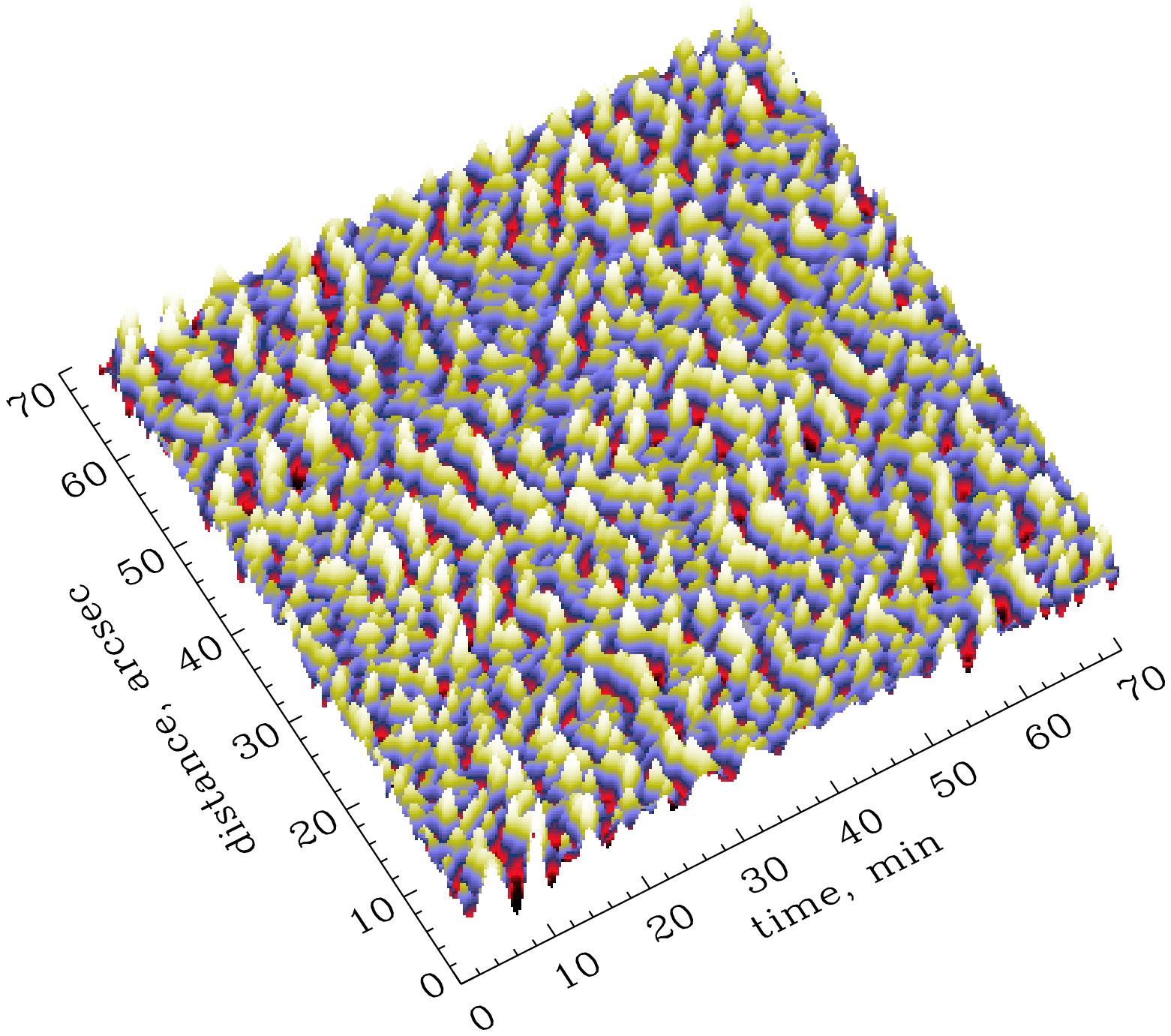}
\caption{Convective (left panel) and oscillatory (right panel)
components of the velocity field corresponding to heights about
650 km as a function of the slit position and time. }
\label{fig:filtered}
\end{figure*}

We used 55 reference widths of the average \BaII\ 4554 \AA\ line
intensity profile
$\Delta\lambda_W=\overline{\Delta\lambda_r}-\overline{\Delta\lambda_b}$,
where $\overline{\Delta\lambda_r}$ and
$\overline{\Delta\lambda_b}$ are the wavelengths of the red and
blue wings, respectively. For each individual \BaII\ profile at
each spatial position $x$ (varying from 1 to 730) and time moment
$t$ (varying from 1 to 600), we then found intensity variations
$\delta I(t,x,W)$ and velocities $V(t,x,W)$ at the corresponding
55 reference widths. We slightly modified the standard
$\lambda$-meter procedure as described below, keeping in mind that
the formation heights of the blue and red wing intensity points of
the line profile are, in general, not the same because of the
Doppler velocity shift
\citep[see \eg][Paper I]{Shchukina+etal2009}.
The velocities were calculated separately in the red and blue
wings of the \BaII\ line in the following way:
\begin{eqnarray}
\label{velo}V_r(t,x,W) & = &  \delta V_r(t,x,W)+\bar{V}(W)  \\
\nonumber V_b(t,x,W) & = &  \delta V_b(t,x,W)+\bar{V}(W)\,
\end{eqnarray}
where
\begin{eqnarray}
\label{vari}
 \delta V_r(t,x,W) & = & (c/\lambda)\cdot{\delta\lambda_r} \\
\nonumber
 \delta V_b(t,x,W) & = & (c/\lambda)\cdot{\delta\lambda_b} \,.
\end{eqnarray}
In these equations, $\bar{V}(W)$ is the  bisector of average
profile due to to nonthermal velocity field
\begin{eqnarray}
\label{bisec} \bar{V}(W)  & = &
(c/\lambda)\cdot(\overline{\Delta\lambda_r}+\overline{\Delta\lambda_b})/2
-
V_{\rm HFSI} \,, 
\end{eqnarray}
where $V_{\rm HFSI}$ is the line bisector caused by the hyperfine
structure (HFS) and isotopic shift, $c$ velocity of light and,
$\lambda$ the line wavelength. Furthermore, in eq.~(\ref{vari})
$\delta\lambda_{b,r}=\Delta\lambda_{b,r}
-\overline{\Delta\lambda_{b,r}} $ denotes the wavelength shift of
the blue (red) intensity point  belonging to a certain spectral
width relative to the corresponding blue (red) point of the
average profile.

We derived the bisector $V_{\rm HFSI}$ using the \BaII\ 4554 \AA\
line profile calculated in the MACKKL model
\citep[see \eg][Paper I]{Shchukina+etal2009}.
Figure~\ref{fig:asym} shows that the bisector $V_{\rm HFSI}$ is
negative (between $-0.3$ and $-0.2$ km/s), while the nonthermal
velocity bisector $\bar{V}(W)$ of the average observed profile
changes sign ranging from $-0.1$~km/s to $\approx 0.25$~km/s. The
velocity values were corrected for the displacement due to the
Earth rotation and revolution.
Note that through this paper negative velocities mean downflows
and positive velocities mean upflows. The only exception is
Fig.~\ref{fig:asym} where, in agreement with Fig.~2 of Paper I, we
assume negative velocity direction to be toward the observer,
corresponding to an upflow.

In a similar way, we calculated intensity variations (contrast)
$\delta I(t,x,W)$  relative to the intensity $\bar{I}(W)$ of the
\BaII\ line profile averaged over space and time:
\begin{eqnarray}
\delta I(t,x,W) & = &
(I(t,x,W)-\bar{I}(W))/{\bar{I}(W)} \,. 
\end{eqnarray}

A through analysis of the Doppler diagnostic techniques by
\citet[][Paper I]{Shchukina+etal2009}  has shown that the $V_b$
and $V_r$ velocities calculated from the line profile of the
\BaII\ 4554 \AA\ line are, in fact, very close to each other.
According to their analysis, both $V_b$ and $V_r$ correspond
better to heights of formation $H_b$ of the blue wing intensities,
rather than to $H_r$ of  the red wing intensities. Guided by the
results of Paper I we use the average values $V=(V_b+V_r)/2$
throughout the present paper  and ascribe their origin to the
average heights of formation of the blue wing intensities.

Variations of intensity $\delta I(t,x,W)$ and velocities
$V_r(t,x,W)$ and $V_b(t,x,W)$ are mainly caused by oscillatory and
convective motions. By convective motions we mean the motions of
individual granules and intergranular lanes. To separate
oscillatory and convective components we performed filtering in
the Fourier domain based on the diagnostic $k-\omega$ diagram.
According to this diagram, we limited the wave motions in temporal
frequency between $\omega=$1.8 and 5.7 mHz and convective motions
by frequencies below $\omega=2.2$ mHz. Additional filtering was
performed for convective motions in the spatial frequency, leaving
only variations with $k$ less than 0.18 Mm$^{-1}$. Similar
filtering was applied in \citet[][]{Khomenko+etal2001} and
\citet{Kostik+Khomenko2007}. As the frequency domains overlap, the
wave motions can, in principle, contribute to convective motions.
The maximum contribution of the wave motions at $\omega=3.3$ mHz
reaches 4.2\%. We checked that decreasing this contribution to
3.1\% (more narrow filter) or increasing it to 20\% (more wide
filter) produces only an insignificant influence on our results.
As an example we show in Fig.~\ref{fig:filtered} the convective
(left) and oscillatory (right) components of the velocity
corresponding to heights around 650 km as a function of space and
time. The vertical axis represents the velocity amplitudes. The
different temporal and spatial behavior of the both components is
evident from the figure. In the rest of the paper we focus our
analysis on the convective component of the velocity and intensity
variations. In addition, we only select a part of the time series
(of about 200 exposures) when the seeing conditions were the best;
i.e., the estimated spatial resolution was not worse than
0.\arcsec4.

\begin{figure}
\centering
\includegraphics[width=9cm]{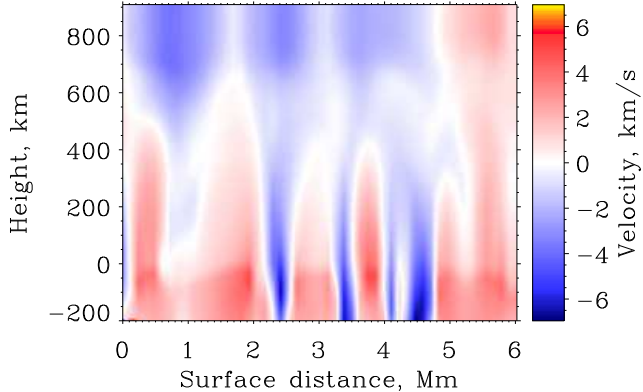}
\caption{Vertical velocities along one of the slices in the 3D
model. Negative velocities correspond to intergranules (downflow,
blue color), while positive velocities correspond to granules
(upflow, red color).} \label{fig:Vz}
\end{figure}

\section{Formation heights along the observed  \BaII\ line profiles}

The formation heights of the 55 reference widths of the \BaII\
4554 \AA\ intensity profiles were obtained from NLTE calculations
in a 3D hydrodynamic model atmosphere by
\citep{Asplund+Nordlund+Trampedach+AllendePrieto+Stein2000b}.
Detailed description of the Barium atomic model, NLTE
calculations, and the 3D hydrodynamical snapshot is given in
\citet{Olshevsky+etal2007} and \citet[][Paper
I]{Shchukina+etal2009}. In this section we  present additional
information that can be useful for understanding our results.

Figure~\ref{fig:Vz} visualizes vertical variations in the velocity
along one of the slices of this snapshot. The geometrical height
is the same in all the points of the snapshot, but the optical
depth scale is fluctuating from column to column due to horizontal
inhomogeneities of the thermodynamic quantities.
The zero point of geometrical height $H=0$ is assigned by
averaging over all grid points of the snapshot the locations where
the continuum optical depth $\tau_{5000}=1$
\citep[see][]{Asplund+Nordlund+Trampedach+AllendePrieto+Stein2000b}.

The vertical extent of the snapshot was optimized to reach high
enough layers (till 900~km) to allow reliable calculation of the
cores of strong lines \citep[see][]
{Asplund+Nordlund+Trampedach+AllendePrieto+Stein2000b,
Stein+Nordlund1998, Nordlund+Dravins1990}. Taking into account
that the core of the \BaII\  4554 \AA\ line is formed more than
100 km below the upper boundary \citep[][Paper
I]{Shchukina+etal2009}, one can expect the effects of the upper
boundary conditions to be negligible for spectral synthesis of
this line. An excellent agreement of the spatially average
synthetic profile and observed profile  from the Li\'ege atlas
\citep{Liege1973}, as well as profile obtained from observations
at the VTT, supports this conclusion \citep[see Fig.~4 in][Paper
I]{Shchukina+etal2009}.

The solar disk-center profiles of the \BaII\ 4554 \AA\ line were
calculated by means of 1D multilevel NLTE radiative transfer
employing  a set 1D models from individual grid points of the 3D
snapshot as input. Such an approach is known in the literature as
1.5D approximation.

For each grid point (1D model), we calculated formation heights of
the blue wing intensities $H_b$ corresponding to 55 reference
widths $\Delta\lambda_W$ of the \BaII\ profile. Following
\citet[][Paper I]{Shchukina+etal2009},  we used the concept of
Eddington-Barbier heights of line formation; i.e., we evaluated
heights where the line optical depth at a given wavelength point
${\Delta \lambda}$ is equal to unity: $\tau({\Delta \lambda}) =1$.
We then assumed that the information on vertical velocity and
intensity at wavelength point ${\Delta \lambda}$ comes from this
height. We repeated the calculations of $H_b$ heights for the all
grid points of the 3D snapshot and then computed an average value
$\langle H \rangle$ for each of the 55 reference widths
$\Delta\lambda_W$. These formation heights were ascribed to the
corresponding widths of the observed intensity profiles. Thus, the
formation heights calculated this way are model-dependent.

According to \citet[][Paper I]{Shchukina+etal2009}, the mean
formation heights $\langle H \rangle$ of each section of the
\BaII\ 4554 \AA\ line profile have a well-pronounced dependence on
the spectral width of this section. Thus, the intensity and
velocity information coming from each width level
$\Delta\lambda_W$ of the observed profile roughly corresponds to a
different height in the atmosphere. The selected widths $\Delta
{\lambda}_W$ of observed profiles  vary between 280~m\AA\ and
30~m\AA. The corresponding range of the mean formation heights
lies between $\sim -25$~km and $\sim 650$~km.

The lowest height level corresponds to the width
$\Delta\lambda_W=280$~m\AA,  and its average formation height is
$\langle H \rangle=-25$~km. We used the intensity variations at
this level as a criterion for separating granular and
intergranular regions. Taking into account that this height is
expected to be close  to the formation height of the continuum
intensity, we call ``granules'' those features whose intensity
$I(t,x,W)$ at $\Delta\lambda_W=280$~m\AA\ ($\langle H
\rangle=-25$~km) was larger than the average $\bar{I}(W)$. The
classical definition of ``granule'' and ``intergranule'' can only
be applied to the continuum formation heights. Higher in the
atmosphere, this definition does not make sense.

%
\begin{figure}
\centering
\includegraphics[width=9cm]{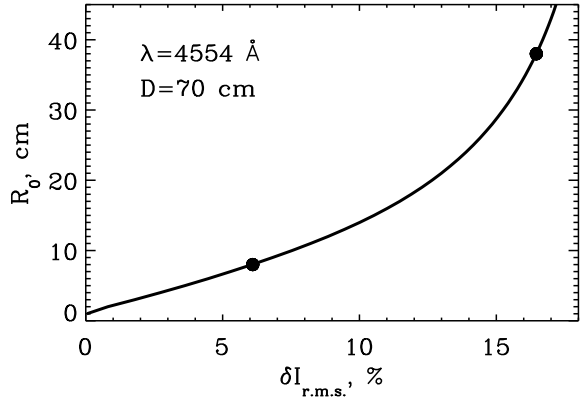}
\caption{Root-mean-square continuum contrast $\delta I_{\rm
r.m.s.}$ as a function of Fried's parameter $R_0$ for wavelength
$\lambda=4554$~\AA\ and for telescope diameter $D=70$~cm. The
$\delta I_{\rm r.m.s.}$ curve was calculated using 3D radiative
transfer formal solution for continuum intensity obtained by
\citet{trujillo:shchu:2008} in the 3D HD model. Filled circles
correspond $R_0=8$~cm ($\delta I_{\rm r.m.s.}=6.1$~\%) and
$R_0=38$~cm ($\delta I_{\rm r.m.s.}=16.5$~\%). }
\label{fig:contrasts_3D}
\end{figure}

\subsection{Spatial smearing and determination of Fried's
parameter} \label{sec:fried}

To directly compare between the synthetic and observed spectra, we
performed a convolution of the synthetic profiles with a
corresponding  modulation transfer function (MTF) representing the
action of the 70-cm telescope and  Earth's atmosphere \citep[see
\eg ][]{Khomenko+etal2005, Shchukina+etal2009}. In such an
interpretation the quality of the resulting image depends on the
telescope diameter $D$ and Fried's parameter $R_0$
\citep{fried:1966}.
To perform the smearing, the original 2D intensity maps at each
individual wavelength were Fourier-transformed and multiplied by
MTF calculated for known telescope diameter (70 cm, VTT) and
Fried's parameter $R_0$. The corresponding MTFs are given in
\citet{Shchukina+etal2009}. An inverse Fourier transform was
performed to obtain the smeared images, i.e., the images affected
by the diffraction on the telescope aperture and the seeing
effects.

\citet{ric:etal:1981} proposed determining $R_0$ from the observed
r.m.s. continuum contrast of solar granulation $\delta I_{\rm
r.m.s.}$ Following their idea, \citet{Shchukina+etal2009}
calculated theoretical dependences between $\delta I_{\rm r.m.s.}$
and the Fried's parameter $R_0$ at different wavelengths.
These  authors used  3D radiative transfer solution for the solar
continuum intensity obtained by \citet{trujillo:shchu:2008} in the
same snapshot of the 3D HD model. We applied their theoretical
calibration curve at wavelength $\lambda=4554$~\AA\ to define
Fried's parameter $R_0$ from our observations (see
Fig.~\ref{fig:contrasts_3D}).
We found that, with the spatial resolution reached during the
observations ($\sim$0.\arcsec7$-$0.\arcsec3), the root-mean-square
continuum contrast $\delta I_{\rm r.m.s.}$ at the wavelength of
the \BaII\ line is between 5.4\% and 6.3\%. Thus, we estimate
Fried's parameter $R_0$ in the range  between 7~cm and 8~cm.


\section{Results of observations and comparison with model
calculations}

According to \citet[][Paper I]{Shchukina+etal2009}, the
$\lambda$-meter technique applied to the  \BaII\ 4554 \AA\ line
profiles recovers velocities at different heights rather well.
Nevertheless, there are three important points that we have to
keep in mind.

First, under seeing conditions corresponding to our observations,
($R_0=7-8$~cm) the velocities derived from the inner wings
($\Delta\lambda_W < 150$~m\AA) of the \BaII\ 4554 \AA\ line
profiles are more reliable than from the outer wings (see Paper I,
Fig.~9). In the former case, the correlation coefficient between
the original and $\lambda$-meter velocities varies between
$0.7-0.8$, while in the latter it becomes lower (about 0.6).

Second, such a parameter as the average formation height should be
used with caution. In reality, the  intensities corresponding to a
single value of $\Delta\lambda_W$ are formed in a wide range of
heights, depending on the model atmosphere. The peak-to-peak
formation height variations can be as much as 200~km in the outer
blue wings and even more ($\sim 500$~km) in the inner blue wings
(see Paper I, Fig. 13).

Third, the absolute values of the $\lambda$-meter velocities are
lower than the original ones. This is not surprising. On the one
hand, we recover velocities from the synthetic profiles degraded
by atmospheric seeing. This should lower the absolute values of
the velocities. On the other hand, the $\lambda$-meter technique
itself leads to smoothing and vertical averaging of the velocities
within the heights where the line intensity is formed.

\begin{figure}
\centering
\includegraphics[width=9cm]{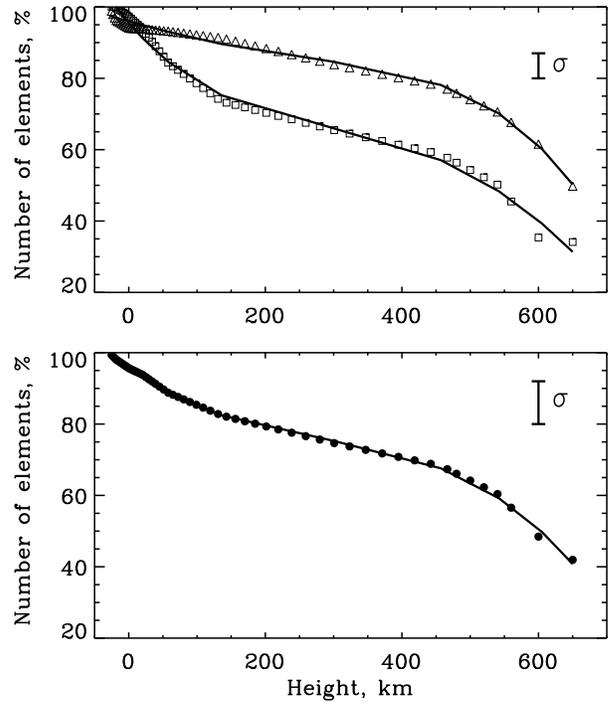}
\caption{Top: fraction of the spatial locations where the
intensity contrast sign (squares) and the velocity direction
(triangles) at a given height (or $\Delta\lambda_W$) changes along
the slit. Bottom: same, but averaged over the velocity and
intensity events. } \label{fig:cinvert}
\end{figure}

\begin{figure}
\centering
\includegraphics[width=9cm]{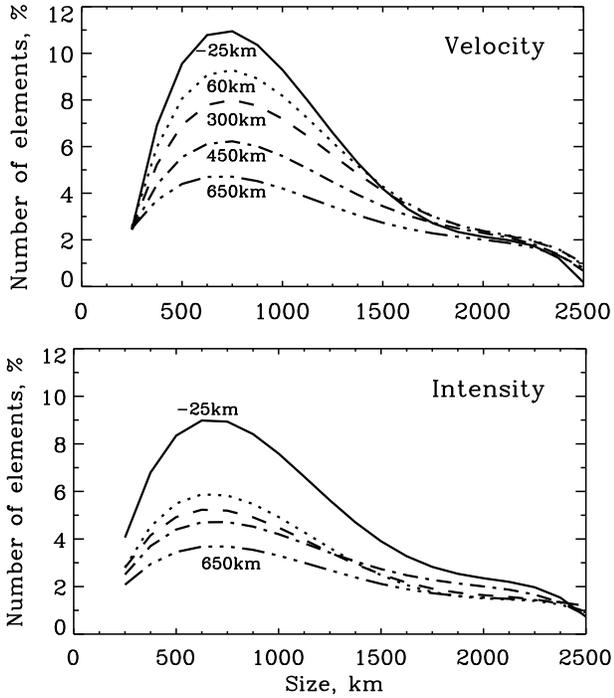}
\caption{The fraction of the convective structures with a given
size at different heights in the atmosphere. The curves (from the
upper to the lower) correspond to average heights $ \langle H
\rangle = -25$, 60, 300, 450, and 650 km. Top panel: size
calculated from the velocity variations; bottom panel: size
calculated from the intensity variations.} \label{fig:sizes}
\end{figure}

Taking these limitations into account, we discuss below in this
section the results of application of $\lambda$-meter technique to
our observations.
\citet{Kostyk+Shchukina2004} have shown that the granular motions
in the lower photosphere preserve their column structure
up to height of 490 km. 
We can verify this result because the \BaII\ line allows us to
investigate the properties of the granulation at greater heights
up to 650 km.
Using 200 spectral images taken during the best seeing conditions
of the series, we calculated the number of ``convective columns''
at each of 55 $\Delta\lambda_W$ positions. This number was defined
as a number of contrast sign changes (or, independently, the
velocity sign changes) along the slit at each height. We normalize
the number of these columns (independently for velocity and
intensity) to the total number at the lowest layer. The results
are presented in Fig.\ref{fig:cinvert}. The horizontal axis of the
both panels of this figure gives the average NLTE formation
heights corresponding to 55 values of the profile widths
$\Delta\lambda_W$. The vertical axis of the top panel gives the
percentage of the locations of intensity contrast (squares) and
the velocity direction (triangles) changes along the surface,
calculated with respect to the lowest layer. The bottom panel of
the figure gives similar information but averaged over the
velocity and intensity.
As expected, the number of the locations where the contrast sign
or the direction of motion changes decreases with height, both for
velocity and intensity events (top panel of
Fig.~\ref{fig:cinvert}). There are about 20\% fewer events for
intensity at all heights.
However, an unexpected result (at least for us) was that about a
half of the convective structures (40\%) registered in the
continuum are maintained with height as high as at 650 km.


Next we investigated how the size of the convective structures
changes with height. We define the size as a simple difference in
spatial coordinate between the successive changes in the sign of
the contrast or in sign of the velocity. The results are shown in
Fig.~\ref{fig:sizes}. The horizontal axis of the figure gives the
size of the convective structures, and the vertical axis gives the
fraction of the structures of a given size in percent. The curves
are normalized to the corresponding number of elements found at
the lowest layer from the velocity variations.
The upper curves in both panels in Fig.~\ref{fig:sizes} correspond
to the average height of $ \langle H \rangle =-25$~km and the
lower curves to the height of 650~km. The maximum of both velocity
and intensity distributions appears at 650--700~km granular size.
The number of structures with this size  decreases  with height
from about 11\% at $-25$~km to $\sim 4\%$ at 650~km (as defined by
velocity variations) and from $\sim 9\%$ to $\sim 3\%$ (as defined
by the intensity variations). Convective structures that have
larger sizes (about 1500-1700~km) at the height of continuum
formation almost always reach the upper layers. Their fraction
with height is almost constant. It means that the average size of
convective elements gets larger with height.
Here we need to give additional explanations. We did not follow
the behavior of each individual structure with height. It means
that, most probably, not all the large-size convective structures
with sizes above 1500~km at $ \langle H \rangle =-25$~km reach a
height of 650~km. It can instead be a result of merging of the
convective structures with smaller sizes. In either case, we can
conclude that, in the height range 0--650~km in the solar
atmosphere, the fraction of the convective structures with sizes
1500--2500~km is almost constant.

\begin{figure}
\centering
\includegraphics[width=9cm]{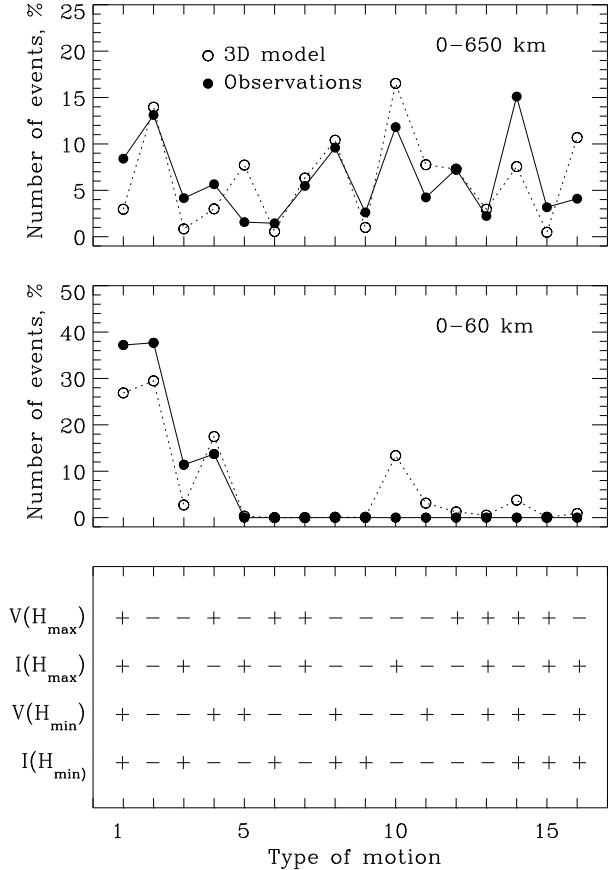}
\caption{Convective motions in the solar atmosphere according to
16-column model. Upper panel shows the number of cases
corresponding to each of the 16 types of motions (as defined at
the bottom panel) between 0 and  650 km. Middle panel gives the
same but for heights between 0 and 60 km. Solid line:
observations. Dotted line: 3D HD model. } \label{fig:columns}
\end{figure}

How do individual properties of the convective elements change
with height? We assume that the convective elements observed at
the uppermost layer at 650 km can either preserve their contrast
and velocity direction from the continuum formation height or
change them by the opposite. This way we define 16 types of
motion.
These types of motion are outlined in the lower panel of
Fig.~\ref{fig:columns}. The ``$+$'' sign stands for convective
elements moving upwards or those whose contrast is above the
average. The ``$-$'' sign stands for those elements moving
downwards or having the contrast below the average.
All 16 types of motions are indeed present in the solar
atmosphere.
The upper panel of Fig.~\ref{fig:columns} gives the number of
cases corresponding to each of the 16 types of motion between 0
and 650 km heights, in percent (solid line). The number of cases
changes from the lowest value of 1.4\% (case number 6; downward
moving hot material in the continuum and upward-moving cold
material at 650 km) to the highest value of 15.1\% (case number
14; upward-moving hot material in the continuum and upward-moving
cold material at 650 km).
Only 21.4\% of the elements show the pure convective character of
motion (\ie\ hot material rises, cold material sinks). The latter
number contains the sum of the cases 1 and 2 ($8.3+13.1$). This
number can be even lower if the convective motions change their
direction of motion or sign of contrast several times between 0
and 650 km. In contrast, in some 9.9\% of the cases, the cold
material rises and the hot material sinks (cases 3 and 4). The
remaining 68.7\% of the convective elements either change their
sign of contrast (31.7\%) or direction of motion (18.8\%) or both
(18.2\%) between 0 and 650 km.

The middle panel of Fig.~\ref{fig:columns} gives the properties of
the convective motions in the low photosphere close to the
continuum formation level (solid line) between 0 and 60 km. Most
of the convective elements preserve their properties within this
low height range. However, even in this case only 75\% of the
elements follow the classical convection behavior, while about
15\% of the relatively cold material moves upwards and 10\% of the
relatively hot material moves downwards.

Similar analysis was also performed with synthetic line profiles
(see dotted curves in the upper and middle panels of
Fig.~\ref{fig:columns}). It is evident that the 3D hydrodynamical
model of
\citet{Asplund+Nordlund+Trampedach+AllendePrieto+Stein2000b}
describes many properties of the observed variations in intensity
and velocity, both qualitatively and quantitatively. Only the case
of motion number 10 has a highest percentage in the simulations
(15\%) at heights 0--60 km compared to its absence in the
observations (middle panel).

The results shown in Fig.~\ref{fig:columns} are unlikely to be
produced by noise in the observations or the influence of the
upper boundary conditions in the simulations. By applying the
filter to separate the wave and the convective motions, the
high-frequency noise component is completely excluded from the
observed variations. As for the simulations, the \BaII\ line is
formed about 100 below the upper boundary of the simulation
domain, so its influence should be negligible.

\begin{figure*}
\centering
\includegraphics[width=16cm]{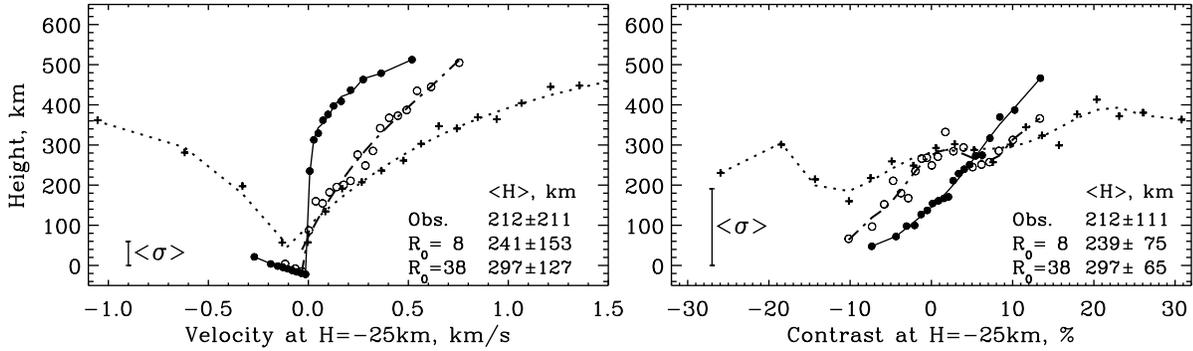}
\caption{Height of the velocity sign reversal as a function of the
velocity (left) and contrast (right) at the average height
$\langle H \rangle =-25$~km. Thick solid lines with filled circles
result from observations, dash-dotted lines with open circles and
dotted lines with crosses result from synthetic profiles convolved
with an MTF function representing the action of the 70-cm
telescope and the Earth atmospheric turbulence with Fried's
parameters $R_0=8$~cm and $R_0=38$~cm, respectively. Each symbol
is an average over bin with an equal number of data points.
Standard deviation $\langle \sigma \rangle$ averaged over all bins
is shown in each panel. The mean heights of the velocity sign
reversal are given in the lower right corner.}
\label{fig:reversal_6}
\end{figure*}

\begin{figure*}
\centering
\includegraphics[width=16cm]{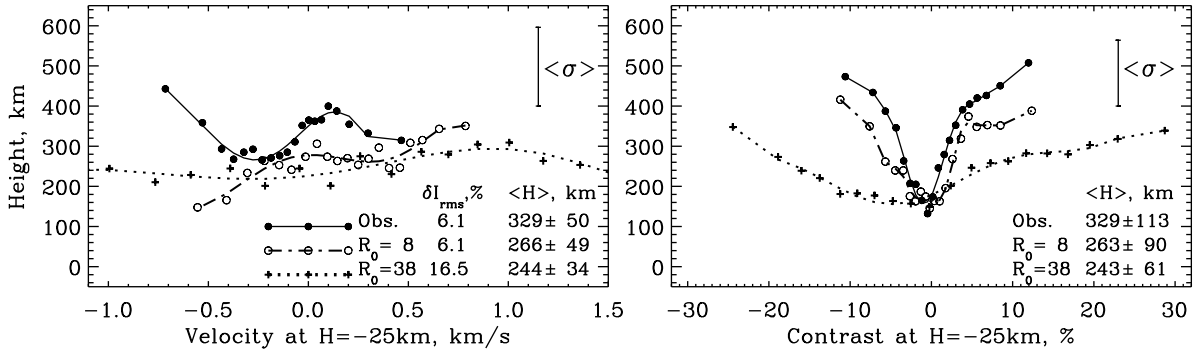}
\caption{Height of the contrast sign reversal as a function of the
velocity (left) and contrast (right) at the average height
$\langle H \rangle =-25$~km. The format of the figure is the same
for as Fig.~\ref{fig:reversal_6}.} \label{fig:reversal_7}
\end{figure*}

\begin{figure*}
\centering
\includegraphics[width=16cm]{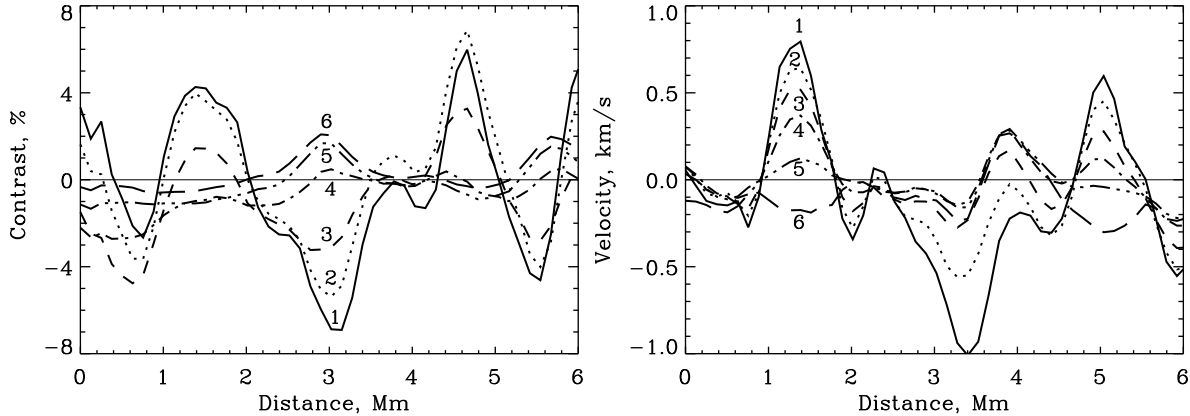}
\caption{Left: Observed variations of the granulation contrast
along the solar surface at several heights. The curves numbered
from 1 to 6 correspond to the average heights:  $\langle H \rangle
= $ 0, 60, 300, 450, 600, and 650 km. Right: same for the
velocities. } \label{fig:vznak}
\end{figure*}

\begin{figure*}
\centering
\includegraphics[width=16cm]{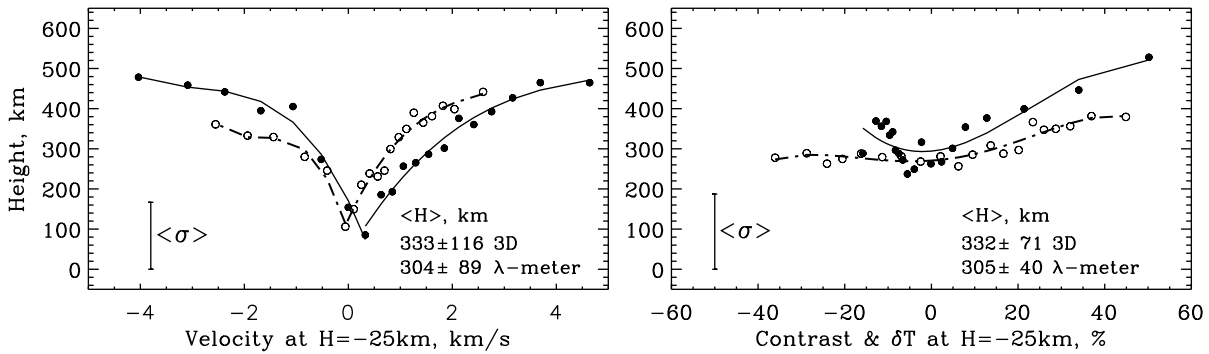}
\caption{Height of the velocity sign reversal as a function of the
velocity (left) and  as a function of contrast or temperature
variations $\delta T$ (right) at $H = -25$~km. Thick solid lines
with filled circles result from 3D model, and dashed lines with
open circles result from synthetic profiles for the case of no
smearing. In the case of the synthetic profiles, the velocities
and contrasts given on the horizontal axes are the values measured
from the profiles at $\Delta\Lambda_W=280$~m\AA\, corresponding to
the average height $\langle H \rangle =-25$~km. In the case of the
3D model, we take the velocities and the temperature variations at
the single layer $H=-25$ km of each model grid point. }
\label{fig:reversal_63d}
\end{figure*}

\begin{figure*}
\centering
\includegraphics[width=16cm]{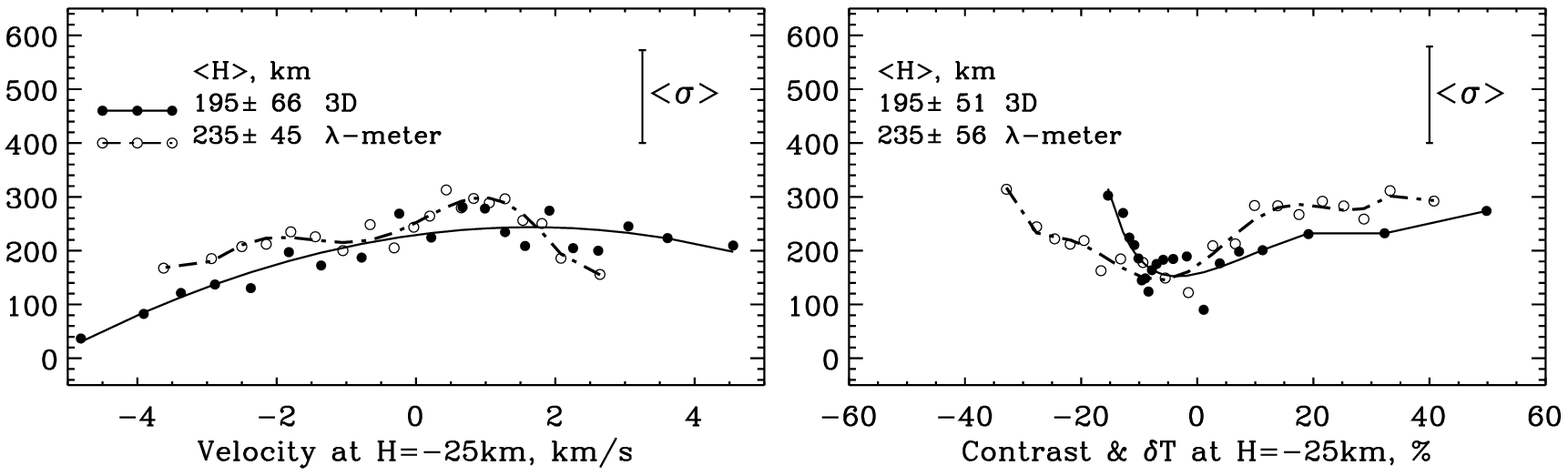}
\caption{Height of the contrast sign reversal as a function of the
velocity (left) and AS A FUNCTION OF contrast or temperature
variations $\delta T$ (right) at $H =-25$~km. The format of the
figure is the same as for Fig.~\ref{fig:reversal_63d}.}
\label{fig:reversal_73d}
\end{figure*}

%
According to the conclusions derived from Fig.~\ref{fig:columns},
the convective elements not only change their sign of contrast
with height but also change the direction of the velocity. The
latter property has not received much attention in the literature,
except the work of \citet{Kostyk+Shchukina2004}. What are the
typical heights where the sign reversal of the convective
intensity and velocity takes place? From the first look at our
observations, we find that there is a big scatter in such heights,
from 50 to 650 km. However, we find that the cause of such a big
scatter is not in systematic or random errors. This scatter comes
from the intrinsic properties of convective elements, \ie\ their
contrast and velocity. This conclusion is illustrated in
Figs.~\ref{fig:reversal_6} and \ref{fig:reversal_7} for
observations with the best r.m.s. continuum contrast $\delta
I_{\rm r.m.s.}\approx 6$~\% (solid curves with filled circles).
The solid curve in  the left panel of Fig.~\ref{fig:reversal_6}
shows that the height of the velocity sign reversal depends on the
strength of the granular velocities at the lowest
level\footnote{As discussed in Sect. 4, the heights are defined
with respect to the average location of $\tau_{\rm 5000}=1$ over
all grid points of the 3D snapshot. The velocities and intensities
at $\langle H \rangle=-25$, as given on the horizontal axis of
Figs.~8 and 9, are measured from the observed profiles at
$\Delta\lambda_W \sim 280$.}.
The higher is the velocity of granules, the higher in the
atmosphere the velocity sign reversal takes place. The curve is
asymmetric with respect to zero velocity value. Given the same
absolute value of the velocity, the downward-moving convective
elements change their sign of velocity at substantially lower
heights than the upward moving elements. The average value of the
height of the velocity sign reversal is 212$\pm$211~km.

The right panel of  Fig.~\ref{fig:reversal_6} gives the velocity
sign reversal height as a function of contrast corresponding  the
most bottom layer  ($\langle H \rangle =-25$~km). Similar to the
above, the higher the contrast of granules in the continuum, the
greater the height of the velocity sign reversal. There is again
an asymmetry in the granulation properties. The bright granules
change their sign of velocity at much greater heights than the
dark intergranular lanes. The difference can be as much as 300 km
for the features with the continuum contrast of $\pm$7.5\%.

Figure~\ref{fig:reversal_7} illustrates the dependence of the
height of contrast sign reversal on the velocity (left panel) and
intensity (right panel) at $\langle H \rangle =-25$~km. The left
panel of this figure shows that, the lower the downward convective
velocity at bottom level, the greater the height of the contrast
sign reversal of these elements. This is what expected considering
that the observed correlation between convective velocity and
intensity in the deep photospheric layers is rather high
\citep[\eg][]{Kostyk+Shchukina2004}. On average, the contrast sign
reversal takes place at $ \langle H \rangle  = 329 \pm 50$ km,
which is higher than the velocity sign reversal by $\sim 100$~km.
According to the right panel of Fig.~\ref{fig:reversal_7}, the
contrast sign reversal occurs approximately at the same height
both for granules and intergranules, as defined by their continuum
intensity. The higher the absolute value of the continuum
intensity, the higher the reversal occurs in the atmosphere.

In general, the agreement between the observed dependences and the
synthetic ones calculated for the same seeing ($R_0=8$~cm) is
fairy satisfactory.
However, the agreement is worse for results presented in the left
panel of Fig.~\ref{fig:reversal_7}, where the contrast sign
reversal height of the elements with downward granular velocities
calculated from 3D model are considerably lower. The reason for
that lies in the poor correlation between the velocity and
contrast at the bottom layer  $\langle H \rangle =-25$~km of the
3D model. In the 3D model, the correlation coefficient is rather
low ($\sim 0.35$), while it is appreciably higher, ($\sim 0.65$)
in observations.
These values of the correlation coefficients agree with the
results obtained earlier from three iron lines by
\citet{Kostyk+Shchukina2004}.

Figure~\ref{fig:vznak} gives additional illustration of the
granular velocity and intensity behavior with height. It shows
several examples of the observed  horizontal variations of the
contrast (left) and velocity (right) along the solar surface at
different heights. The curves demonstrate clearly that the
amplitudes of the variations becomes progressively weaker with
height and the sign reversal takes place. Such a  reversal is
especially evident at spatial locations around 1.5, 3, and 5 Mm.


The dependences discussed above from
Figs.~\ref{fig:reversal_6}--\ref{fig:vznak} were obtained from the
ground-based observations under good seeing conditions. To what
extent are such dependences sensitive to the degrading effects
caused by Earth's atmospheric turbulence? To answer this question,
we applied the same analysis to the set of synthetic profile
smeared with two values of the Fried's parameters $R_0$. The
results for $R_0=8$~cm (resolution similar to our observations)
are shown by open circles with dashed curve in
Figs.~\ref{fig:reversal_6} and \ref{fig:reversal_7}. The results
for $R_0=38$~cm (excellent spatial resolution rarely reached in
ground-based observations) are shown by crosses and dotted curve
in the same figures. We considered also the case of the profiles
with the numerical spatial resolution of the 3D model when the
smearing due to seeing  and instrumental effects is absent. The
latter case is displayed in Figs.~\ref{fig:reversal_63d} and
\ref{fig:reversal_73d} (dash-dotted curves with open circles).

The results obtained with the synthetic profiles clearly
demonstrate that the observed dependences are rather sensitive to
the seeing conditions. According to \citet[][Paper
I]{Shchukina+etal2009}, the spatial smearing reduces the absolute
values of the $\lambda$-meter velocities, particularly of the
strong downward ones. This explains the difference in the
continuum velocity scales on the horizontal axis of
Figs.~\ref{fig:reversal_6}, \ref{fig:reversal_7},
\ref{fig:reversal_63d}, and \ref{fig:reversal_73d} (left panels)
between the cases with $R_0=8$, $R_0=38$, and the original
numerical resolution. In the case with $R_0=38$ and in the case of
the original resolution, the curves have rather symmetric behavior
with respect to the zero velocity point with nearly twice smaller
number of the reversal events above the downflowing areas. Under
the resolution similar to our observations ($R_0=8$~cm), the
information on the velocity sign reversal above the downflowing
areas is lost.
The contrast of granulation also decreases with decreasing the
parameter $R_0$. As a result, all dependences on the granulation
contrast  become smoother and less pronounced under better seeing
conditions.
As follows from Figs.~\ref{fig:reversal_6},~\ref{fig:reversal_7}
and Figs.~\ref{fig:reversal_63d},~\ref{fig:reversal_73d}, the
average height of the velocity sign reversal increases with
improving spatial resolution, whereas the average height of the
contrast sign reversal becomes smaller. The difference can reach
about 100~km between the case with $R_0=8$ cm and the case with
original numerical spatial resolution.

The evidence presented above gives us confidence that our results
on the velocity and contrast sign reversal are not artifacts
caused by specific analysis procedure. Additional evidence is
presented in Figs.~\ref{fig:reversal_63d},~\ref{fig:reversal_73d}
(solid curves with filled circles). These data illustrate the
velocity sign reversal heights and the temperature variation sign
reversal heights measured directly from the 3D model snapshot. The
temperature variations ($\delta T$) are defined relative to the
average temperature ($\bar{T}$) over the snapshot at each
particular height, \ie\ $\delta T =(T-\bar{T})/\bar{T}$. It is
clear from the figure that the dependences measured directly from
the 3D model are very similar to the ones obtained after spectral
synthesis and the $\lambda$-meter technique (see dashed-dotted
curved with open circles on the same figure). The only remarkable
difference is that the $\lambda$-meter velocities and contrast
values are systematically lower. The latter is easy to understand
having in mind that the $\lambda$-meter method gives the average
information over a certain height range, thus leading to such a
reduction.

\section{Discussion}

\begin{figure}
\centering
\includegraphics[width=8.5cm]{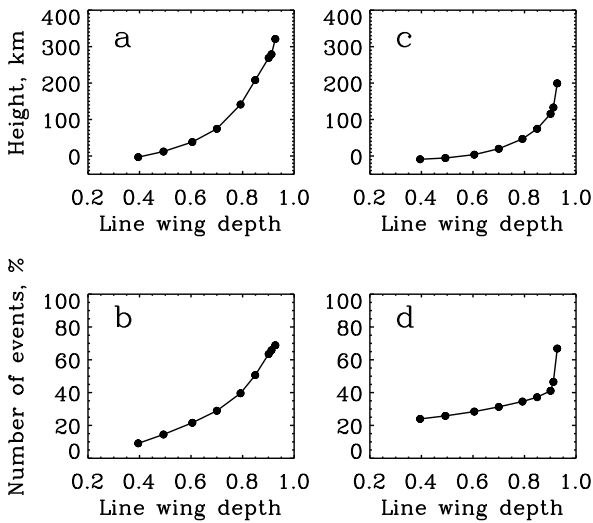}
\caption{{\it a}: average height of the contrast sign reversal as
a function of \BaII\ line depth. {\it b}: fraction of the events
of the contrast sign reversal with height as a function of the
\BaII\ line depth. {\it c} and {\it d}: same for the velocity sign
reversal. } \label{fig:fheight}
\end{figure}

The classical picture of the convection breaks in the solar
photosphere, starting at least from 50 km above the continuum
formation level. The convective elements moving upwards and
downwards not only change their sign of contrast but also their
direction of motion. Often both events take place at the same
time. This process occurs in a wide range of heights up to 650 km.
The height where the contrast sign reversal takes places depends
strongly on the contrast of granules and intergranules at the
continuum formation level. \citet{Stodilka2006} came to the
conclusion that the temperature inversion happens in a wide range
of heights, from 150 to 500 km. In addition, there is a similar
tendency, \ie\ the larger the granule, the higher the temperature
inversion takes place. According to \citet{Puschmann+etal2005},
the temperature inversion occurs for structures with sizes over
1.\arcsec5 at heights above 200 km. Our results confirm these
conclusions. On the other hand, \citet{Stodilka2006} claims that
the temperature reversal events are more often above
intergranules, while according to our results, this happen more
often above granules (see right panel of
Fig.~\ref{fig:reversal_7}). The large scatter in the average
height of the contrast (temperature) reversal according to
observations of different authors (from 60 to 350 km) not only
stems from the observational details (like seeing) and their
analysis, but also from the methods used to calculate the
formation heights of the observed spectral lines, as correctly
pointed out in \citet{Puschmann+etal2005}. We think that the main
reason for this disagreement is the difference in the formation
heights of spectral lines used in different observations. To prove
this statement, consider the calculation presented in
Fig.~\ref{fig:fheight}.
The panel {\it a} of this figure gives the average height of the
contrast sign reversal as a function of the \BaII\ line depth for
the our observations.  It follows that the larger the line depth,
the higher the level of the contrast sign reversal obtained. The
panel {\it b} of Fig.~\ref{fig:fheight} gives the fraction of the
contrast sign reversal events as a function of the line depth.
As one can see, the  stronger lines probe higher atmospheric
regions and may, therefore, detect more sign reversals. Since
different authors use spectral lines of different strengths
(corresponding to  different formation heights of  the central
intensity), it is natural that they obtain different heights of
the sign reversal of solar granulation contrast.
The panels {\it c} and {\it d} of Fig.~\ref{fig:fheight} give
similar results but for the heights where the velocity sign
reversal takes place. The first authors that pointed out the
contrast sign reversal in the solar atmosphere were
\citet{Evans+Catalano1972} and \citet{Holweger+Kneer1989}.
However, not only does the contrast sign reversal take place in
the solar atmosphere, but also the reversal of sign of the
convective velocities with height. On average, it happens at about
200--300 km high. As far as we know, this property of the
convective motions has not ever been studied before.

The comparison of observed dependences with those obtained from
the NLTE synthesis of \BaII\ 4554 \AA\ profiles in the 3D
hydrodynamical model of
\citet{Asplund+Nordlund+Trampedach+AllendePrieto+Stein2000b} have
shown that this model reproduces the observed dependences rather
well except for the dependence of the height of the contrast sign
reversal on the velocity at the bottom layer. We found that all 16
classes of the convective motions are present in the 3D model, and
their statistical behavior is similar to the observed. The
contrast sign reversal and the velocity sign reversal take place
over the whole atmosphere from 50 to 650 km. However, the column
structure of granulation is preserved up to 650 km. Almost 40\% of
all the convective structures detected at the continuum formation
height reach 650 km. In particular, 80\% of the structures larger
than 1500 km in the continuum reach 650 km. Most losses are
experienced by the smallest convective elements with sizes below
500 km, while the fraction of the elements with larger sizes
increases with height.

\section{Conclusions}

We performed statistical analysis of the solar granulation
velocities and intensities extracted by means of a $\lambda$-meter
technique from the high-resolution observations of the quiet solar
disk center in the \BaII\ 4554 \AA\ line. As this line forms high
enough in the atmosphere, it has allowed us to study the
properties of granulation at heights not accessed in the earlier
studies up to 650 km, i.e. the lower chromosphere. The results of
observations were compared with the synthetic data from the 3D
hydrodynamical model of the solar convection by
\citet{Asplund+Nordlund+Trampedach+AllendePrieto+Stein2000b}. We
found similar dependencies in the observations and simulations
smeared to the observational resolution. It gives us confidence in
our results and also proves the good quality of the 3D model. Our
main findings can be summarized as follows
\begin{itemize}
\item About 40\% of the column structure of granulation is
preserved up to heights around 650 km.
\item The average size of granular elements increases with height.
\item The granulation can be described by a 16-column model,
depending on the sign of contrast and velocity direction at the
bottom and top atmospheric levels. All types of motion are found
to indeed be present on the Sun.
\item The contrast sign reversal of granulation takes place at heights around
200--300 km. At the same heights, on average, the velocity sign
reversal also occurs.
\item The heights where the sign reversal
happens strongly depends on the individual contrast and velocity
of granules at the continuum. The larger these parameters, the
higher the sign reversal takes place.
\item The precise values of the heights of velocity and contrast
sign reversal obtained from observations are sensitive to the
spatial resolution within their standard deviation error bars.
\end{itemize}

\begin{acknowledgements}
This research was funded by the Spanish Ministerio de
Educaci{\'o}n y Ciencia through projects AYA2007-63881 and
AYA2007-66502 and by the National Academy of Sciences of Ukraine
through projects 1.4.6/7-233B and 257Kt.
\end{acknowledgements}


{}


\begin{thebibliography}{}
%
\bibitem[\protect\astroncite{Asplund
  et~al.}{2000}]{Asplund+Nordlund+Trampedach+AllendePrieto+Stein2000b}
Asplund, M., Nordlund, {\AA}., Trampedach, R., \mbox {Allende
Prieto}, C.,
  Stein, R.~F. 2000, A\&A, 359, 729

\bibitem[\protect\astroncite{Bendlin \& Volkmer}{1993}]{Bendlin+Volkmer1993}
Bendlin, C., Volkmer, R. 1993, A\&A, 278, 601

\bibitem[\protect\astroncite{Brandt}{2000}]{Brandt2000}
Brandt, P. 2000,
\newblock Encyclopedia of Astronomy and Astrophysics, article 2008, Solar
  Photosphere: Granulation, editor Paul Murdin,
\newblock IOP, Bristol: Institute of Physics Publishing

\bibitem[\protect\astroncite{Delbouille et~al.}{1973}]{Liege1973}
Delbouille, L., Neven, L., Roland, G. 1973,
\newblock Photometric atlas of the solar spectrum from $\lambda 3\ 000$ to
  $\lambda 10\ 000$,
\newblock Institut d'Astrophysique de l'Universit\'{e} de Li\`ege, Li\`{e}ge,
  Belgium

\bibitem[\protect\astroncite{Espagnet
  et~al.}{1995}]{Espagnet+Muller+Roudier+Mein+Mein1995}
Espagnet, O., Muller, R., Roudier, T., Mein, N., Mein, P.,
Malherbe, J.~M.
  1995, A\&AS, 109, 79

\bibitem[\protect\astroncite{Evans \& Catalano}{1972}]{Evans+Catalano1972}
Evans, J.~W., Catalano, C.~P. 1972, Solar Phys., 27, 299

\bibitem[\protect\astroncite{{Fried}}{1966}]{fried:1966}
{Fried}, D.~L. 1966, Journal of the Optical Society of America
(1917-1983), 56,
  1372

\bibitem[\protect\astroncite{Holweger \& Kneer}{1989}]{Holweger+Kneer1989}
Holweger, H., Kneer, F. 1989,
\newblock in R.~J. Rutten, G. Severino (eds.), Solar and Stellar Granulation,
  Vol. 263, Proceedings of the 3rd International Workshop of the Astronomical
  Observatory of Capodimonte, NATO Advanced Science Institutes (ASI) Series C,
  Dordrecht: Kluwer,  173

\bibitem[\protect\astroncite{Janssen \& Gauzzi}{2006}]{Janssen+Gauzzi2006}
Janssen, K., Gauzzi, G. 2006, A\&A, 450, 365

\bibitem[\protect\astroncite{Khomenko et~al.}{2001}]{Khomenko+etal2001}
Khomenko, E., Kostik, R. I.; Shchukina, N. G. 2001, A\&A, 369, 660

\bibitem[\protect\astroncite{Khomenko et~al.}{2005}]{Khomenko+etal2005}
Khomenko, E., Shelyag, S., Solanki, S.~K., V{\"o}gler, A. 2005,
A\&A, 442, 1059

\bibitem[\protect\astroncite{Kiselman}{1994}]{Kiselman1994}
Kiselman, D. 1994, A\&AS, 104, 23

\bibitem[\protect\astroncite{Kneer et~al.}{1980}]{Kneer+etal1980}
Kneer, F.~J., Mattig, W., Nesis, A., Werner, W. 1980, Solar Phys.,
68, 31

\bibitem[\protect\astroncite{Kostik \& Khomenko}{2007}]{Kostik+Khomenko2007}
Kostik, R.~I., Khomenko, E. 2007, A\&A, 476, 341

\bibitem[\protect\astroncite{Kostyk \& Shchukina}{2004}]{Kostyk+Shchukina2004}
Kostyk, R.~I., Shchukina, N.~G. 2004, Astronomy Reports, Vol. 48,
N 9, 769

\bibitem[Maltby et al.(1986)]{malt:etal:1986}
Maltby, ~P., Avrett, E.~H., Carlsson, ~M., Kjeldeseth-Moe, ~O.,
Kuruch, ~R., Loeser, ~R. 1986, \apj, 306, 284

\bibitem[\protect\astroncite{Nordlund \& Dravins}{1990}]{Nordlund+Dravins1990}
Nordlund, \AA, Dravins, D. 1990, A\&A, 228, 155

\bibitem[\protect\astroncite{Olshevsky et~al.}{2008}]{Olshevsky+etal2007}
Olshevsky, V., Shchukina, N.~G., Vasil'eva, I. 2008, Kinematika i
Fizika
  Nebesnich Tel, 24 (3), 198

\bibitem[\protect\astroncite{Puschmann et~al.}{2005}]{Puschmann+etal2005}
Puschmann, K.~G., \mbox{Ruiz Cobo}, B., V\'azquez, M., Bonet,
J.~A.,
  Hanslmeier, A. 2005, A\&A, 441, 1157

\bibitem[\protect\astroncite{Puschmann et~al.}{2003}]{Puschmann+etal2003}
Puschmann, K.~G., V\'azquez, M., Bonet, J.~A., \mbox{Ruiz Cobo},
B.,   Hanslmeier, A. 2003, A\&A, 408, 363

\bibitem[Ricort et al.(1981)]{ric:etal:1981}
Ricort, G., Aime, C., Roddier, C., Borgnino, J. 1981, Sol. Phys.,
69, 223

\bibitem[\protect\astroncite{Salucci et~al.}{1994}]{Salucci+etal1994}
Salucci, G., Bertello, L., Cavallini, F., Ceppatelli, G., Righini,
A. 1994,
  A\&A, 285, 322

\bibitem[\protect\astroncite{Schr{\"{o}}ter
  et~al.}{1985}]{Schroter+Soltau+Wiehr1985}
Schr{\"{o}}ter, E.~H., Soltau, D., Wiehr, E. 1985, Vistas in
Astronon., 28, 519

\bibitem[\protect\astroncite{Shchukina et~al.}{2009}]{Shchukina+etal2009}
Shchukina, N., Olshevsky, V., Khomenko, E. 2009, A\&A, accepted
(Paper I)



\bibitem[\protect\astroncite{Stebbins \& Goode}{1987}]{Stebbins+Goode1987}
Stebbins, R.~T., Goode, P.~R. 1987, Solar Phys., 110, 237

\bibitem[\protect\astroncite{Stein \& Nordlund}{1998}]{Stein+Nordlund1998}
Stein, R.~F., Nordlund, \AA. 1998, ApJ, 499, 914

\bibitem[\protect\astroncite{Stodilka}{2006}]{Stodilka2006}
Stodilka, M. 2006, Kinematika i Fizika Nebesnich Tel, Vol. 22, N3,
173

\bibitem[\protect\astroncite{{S{\"u}tterlin} et~al.}{2001}]{Sutterlin+etal2001}
{S{\"u}tterlin}, P., {Rutten}, R.~J., {Skomorovsky}, V.~I. 2001,
\aap, 378, 251

\bibitem[\protect\astroncite{{Trujillo Bueno} \&
  {Shchukina}}{2009}]{trujillo:shchu:2008}
{Trujillo Bueno}, J., {Shchukina}, N. 2009, ApJ, 694, 1364
%
\end{thebibliography}
\end{document}